\documentclass[reprint,prl,showpacs,superscriptaddress,aps]{revtex4-1}
\usepackage{amsmath,amssymb,graphicx,microtype,color}
\usepackage{hyperref}
\usepackage{bm}
\usepackage{graphicx}
\usepackage{lipsum}
\definecolor{purp}{rgb}{0.4,0.1,0.7}

\usepackage{tikz} 

\renewcommand{\vec}[1]{\mathbf{#1}}

\newcommand{\s}{\text{s}}
\newcommand{\vx}{\vec{x}}
\newcommand{\vy}{\vec{y}}
\newcommand{\vq}{\vec{q}}

\newcommand{\gr}{\bm{\nabla}}
\newcommand{\IN}{\text{in}}
\newcommand{\OUT}{\text{out}}

\newcommand{\figref}[1]{Fig.~\ref{#1}}

\newcommand{\p}{\partial}

\usepackage{setspace}
\begin{document}
\title{Flow-driven branching in a frangible porous medium} 

\author{Nicholas J.\ Derr}
\thanks{N.J.D. and D.C.F. contributed equally to this work.}
\affiliation{John A.\ Paulson School of Engineering and Applied Sciences, Harvard University, Cambridge, MA 02138}
\author{David C.\ Fronk}
\thanks{N.J.D. and D.C.F. contributed equally to this work.}
\affiliation{Department of Organismic and Evolutionary Biology, Harvard University, Cambridge, MA 02138}
\author{Christoph A.\ Weber}
\affiliation{
Max Planck Institute for the Physics of Complex Systems, Dresden, Germany
}
\author{Amala\ Mahadevan}
\affiliation{Woods Hole Oceanographic Institution, Woods Hole, MA 02450}
\author{Chris H.\ Rycroft}
\affiliation{John A.\ Paulson School of Engineering and Applied Sciences, Harvard University, Cambridge, MA 02138}
\affiliation{Computational Research Division, Lawrence Berkeley National Laboratory,
Berkeley, CA 94720}
\author{L.\ Mahadevan}
\affiliation{John A.\ Paulson School of Engineering and Applied Sciences, Harvard University, Cambridge, MA 02138}
\affiliation{Department of Organismic and Evolutionary Biology, Harvard University, Cambridge, MA 02138}
\affiliation{Department of Physics, Harvard University, Cambridge, MA 02138}

\begin{abstract}
Channel formation and branching is widely seen in physical systems where movement of fluid through a porous structure causes the spatiotemporal evolution of the medium in response to the flow, in turn causing flow pathways to evolve. We provide a simple theoretical framework that embodies this feedback mechanism in a multi-phase model for flow through a fragile porous medium with a dynamic permeability. Numerical simulations of the model show the emergence of branched networks whose topology is determined by the geometry of external flow forcing. This allows us to delineate the conditions under which splitting and/or coalescing branched network formation is favored, with potential implications for both understanding and controlling branching in soft frangible media. 
\end{abstract}


\maketitle

Branching patterns, or arborization, in porous media are common in many natural settings that include both living and non-living matter~\cite{fleury2001}. The formation of arborized patterns in physical and chemical systems is driven by a variety of processes all of which involve a combination of erosion, transport and deposition. On the laboratory scale, these processes can involve the chemical dissolution of brittle matrices by a penetrating reactive fluid~\cite{szymczak2011,grodzki2019}, the advective rearrangement of unconsolidated media, dielectric breakdown of conducting media \cite{duxbury1987,zapperi1997}, the formation of fingerlike protrusions in dense granular suspensions~\cite{cerasi1998}, formation of beach rills in natural drainage systems~\cite{schorghofer2004,lobkovsky2008} etc. On planetary scales, melt transport in the mantle arises via branching morphologies that lead to localized  channels of widths up to 100\,m~\cite{mckenzie1984,spiegelman1993,spiegelman2001}, and water-driven erosion and branching in glaciers arises on scales of the order of 10\,m~\cite{hewitt2011}.  In biological systems, the best known arborized systems are vasculatures in plants and animals. These arise through morphogenetic mechanisms involving gradients and physical flows that arrange and rearrange matter through a variety of feedback mechanisms at the cellular, organismal, and societal level \cite{camazine2003,ocko2015}, and examples include slime molds \cite{tero2010}, vascular networks \cite{ronellenfitsch2019}, and nest architectures of social insects \cite{Khuong1303}. 

Models based on porous flow theory~\cite{scheidegger1960,bear1988} are capable of describing flow through these branched networks. However, their formation requires nonlinear models with multiple evolving phase boundaries which are still only partially understood both theoretically and experimentally. Here, we propose a simple model via an effective continuum theory that links flow, permeability and pressure gradients by considering pore-scale grain dislodgement in a relatively brittle structure. Numerical solutions of the resulting governing equations show the emergence of branching morphologies through selective erosion and subsequent flow enhancement. 

\begin{figure}
	\includegraphics[width=\linewidth]{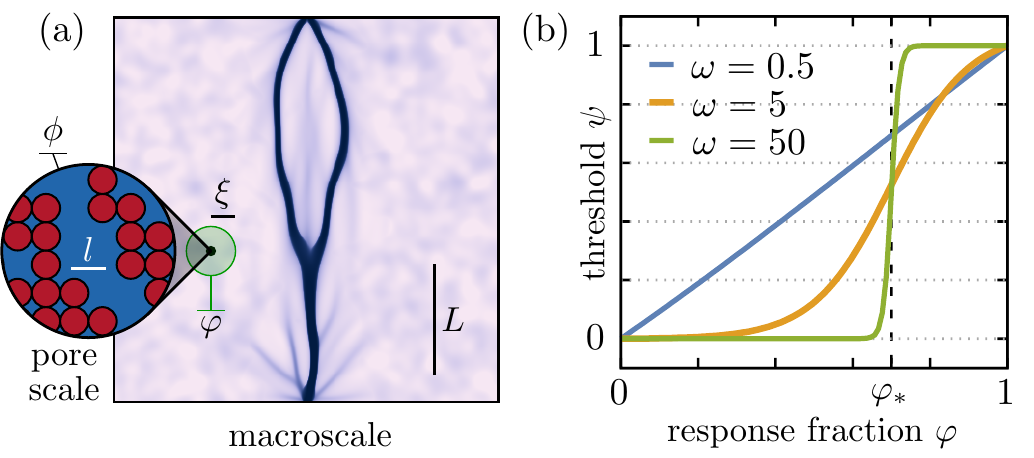}
	\vspace{-1.5em}
	\caption{(Color online) Schematic of model fields, length scales, and erosion criteria. (a) Branched patterns in porous media can emerge on macroscopic lengths $L$ due to interactions at the pore size $l$. In this simulated pattern, obtained by solving \eqref{eq:p} and \eqref{eq:erode}, eroded regions of low solid fraction $\phi(\vx,t)$ are blue (see \figref{fig:branchTopology} for colorbar limits). At a given point on the macroscale (black dot), $\phi$ is  the fraction of a pore-scale integration volume (inset) occupied by rigid grain microstructure (red circles). Fluid-mediated forces on grains induce stresses over a macroscopic region (shaded green circle) characterized by the communication \mbox{length $\xi$}. The response fraction $\varphi$ is the spatial average of $\phi$ throughout this region. (b) The erosion threshold function \smash{$\psi(\varphi) = c_0 \tanh\big(\omega(\varphi-\varphi_*)\big) + c_1 \in [0,1]$} represents resistance to grain dislodgement at  response fraction $\varphi$.}
	\label{fig:microMacro}
\end{figure}

\paragraph{Mathematical model:}  Our starting point is a fluid-filled porous domain $\Omega$ comprised of a rigid grain microstructure with characteristic pore size $l$, as in the \figref{fig:microMacro}(a) inset. The fluid is of viscosity $\eta$ and density $\rho$. On length scales  large compared to the pore size $L \gg l$, we can define macroscopic continuum fields that include the solid fraction $\phi(\vx,t)$ and volumetric fluid flux ${\vq(\vx},t)$ as averages of microscopic quantities \cite{drew1999}. Pressure gradients over macroscopic lengths \mbox{$\Gamma \sim |\gr p|$}  drive motion of the interstitial fluid \mbox{$V \sim |\vq|$}  relative to the pore structure, so that at a scaling level \mbox{$\Gamma \sim \eta V /l^2$}, leading to individual grains feeling forces of magnitude $\Gamma l^3$. When these overcome the attractive forces providing microstructural integrity, grains are dislodged and the local permeability of the medium evolves. If we denote the magnitude of the local network breaking stress by $B(\vx,t)$, which can be heterogeneous, the most general rate law consistent with this mechanism reads
\begin{equation}
	\partial_t \phi = - e_0 \phi f\big(|\gr p|,B\big), \label{eq:erode_gen}
\end{equation}
where $e_0$ is an erosion rate and $f(\alpha,\beta)$ is a nonnegative dimensionless function which vanishes for $\alpha < \beta$. A previous model~\cite{mahadevan2012} accounts for the relative motion of the grains, fluid and the static porous medium via a three-phase model of fluid, immobile solid and mobile grains. Here, we focus on a simpler two-phase model assuming loose grains to be indistinguishable from fluid. 

In terms of a characteristic breaking stress $B_0$ and time scale $\smash{\tau= 1/e_0}$, we can define a characteristic length $L = l(B_0/\eta e_0)$ and pressure gradient magnitude $\Gamma = B_0 / l$. Rescaling our variables and parameters accordingly, and assuming that that the solid is relatively stiff but brittle so that it does not deform, the volumetric fluid flux $\vq$ is well described by Darcy's law,
\begin{equation}
	\vq = -\kappa(\phi)\gr p, \hspace{1em} \kappa(\phi) = \frac{ (1-\phi)^3}{\phi^2}, \label{eq:darcy}
\end{equation}
where the dimensionless permeability $\kappa(\phi)$ is the well-known Carman--Kozeny relation~\cite{scheidegger1960,bear1988}. Assuming the fluid is incompressible, conservation of mass implies
\begin{equation}
	\gr \cdot \vq = -s(\vx,t), \label{eq:mass_conv}
\end{equation}
where $s(\vx,t)$ is the rate at which fluid is depleted due to processes such as bulk reaction or evaporation. By combining the previous two equations, $\vq$ can be eliminated to obtain an elliptic equation for the pressure,
\begin{equation}
	\gr \cdot \big(\kappa(\phi)\gr p\big) = s. \label{eq:p}
\end{equation}
Boundary conditions correspond to specified fluxes \smash{$q_\IN$} and \smash{$q_\OUT$} on boundary regions of inflow \smash{$\p\Omega_\IN$} and outflow \smash{$\p \Omega_\OUT$}  (See Supplementary Information (SI) section SI.1 for details).


\begin{figure}
	\includegraphics[width=\linewidth]{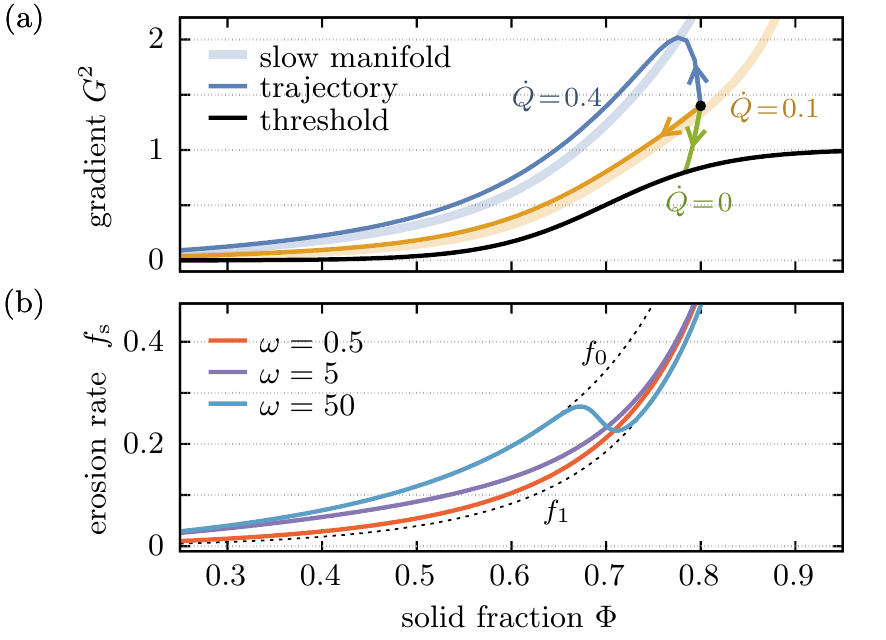}
	\caption{(Color online) Homogeneous model of erosion in the presence of an externally controlled flux \smash{$Q(t)$} defined by \eqref{eq:homogSys}. \mbox{(a) Phase} trajectories through forcing--response space with initial condition \mbox{$(\Phi,G^2) = (0.8,1.4)$} are shown for varying  $\dot Q$. The erosion threshold $\psi(\Phi)$ has $\omega = 8$, $\varphi_*=0.7$.   Constant-flux trajectories reach the threshold quickly, stopping erosion in finite time. For $\dot Q \ne 0$, sustained erosion takes place at long times along the slow manifold $G_{\s}^2(\Phi)$, plotted here as a translucent, thick line. \mbox{(b) The} erosion rate along the manifold, \mbox{$f_{\s} = G_{\s}^2 - \psi$}, is plotted for three thresholds with $\varphi_* = 0.7$ and varying sharpness $\omega$, subject to $\dot Q = 0.1$. Bounds on the rate $f_0 > f_\s >f_1$ are plotted as dotted black lines.} 
	\label{fig:homogSys}
\end{figure}

To close the system, we must relate the erosion rate $f$  to the fields $\phi(\vx,t)$ and $p(\vx,t)$. A minimal analytic form for $f$ with a breaking threshold based on symmetry arguments \cite{mahadevan2012} suggests \mbox{$f = \max\{0,\gr p\cdot \gr p - B^2\}$}, where $B$ is the breaking threshold. The breaking stress itself is a nonlocal function of the solid fraction, depending on the grain density within a region of size $\xi$, a stress communication length which may depend on the porosity. Here, we assume  the following hierarchy of lengths $l  \ll \xi \ll L$, consistent with frangible brittle solids.  In this limit, we introduce a simple erosion threshold $B^2 = \psi\left(\varphi\right)$, defining the response fraction $\smash{\varphi}(\vx,t)$ as the convolution of $\phi(\vx,t)$ with a Gaussian kernel of length scale $\xi$, representing a spatial average of the solid fraction 
as shown in \figref{fig:microMacro}(a) (see SI.2 for details). Thus, the erosion rate law \eqref{eq:erode_gen} becomes
\begin{equation}
	\p_t \phi = - \phi  \max\left\{0,\gr p\cdot\gr p - \psi(\varphi)\right\}. \label{eq:erode}
\end{equation}
For the functional form of the threshold, we consider a sigmoid \smash{$\psi(\varphi) \in [0,1]$}  centered at a critical phase fraction \smash{$\varphi_*$}, where  the behavior is roughly linear over a scale  $\Delta\varphi \sim 1/\omega$, where $\omega$ represents a sharpness parameter as shown in \figref{fig:microMacro}(b). See SI.3 for the exact form. We note that our functional choice satisfies $\psi'\left(\varphi\right) > 0$, i.e. the medium becomes more resistant to erosion at larger \smash{$\varphi$}.  
\begin{figure*}
	\includegraphics[width=0.99\linewidth]{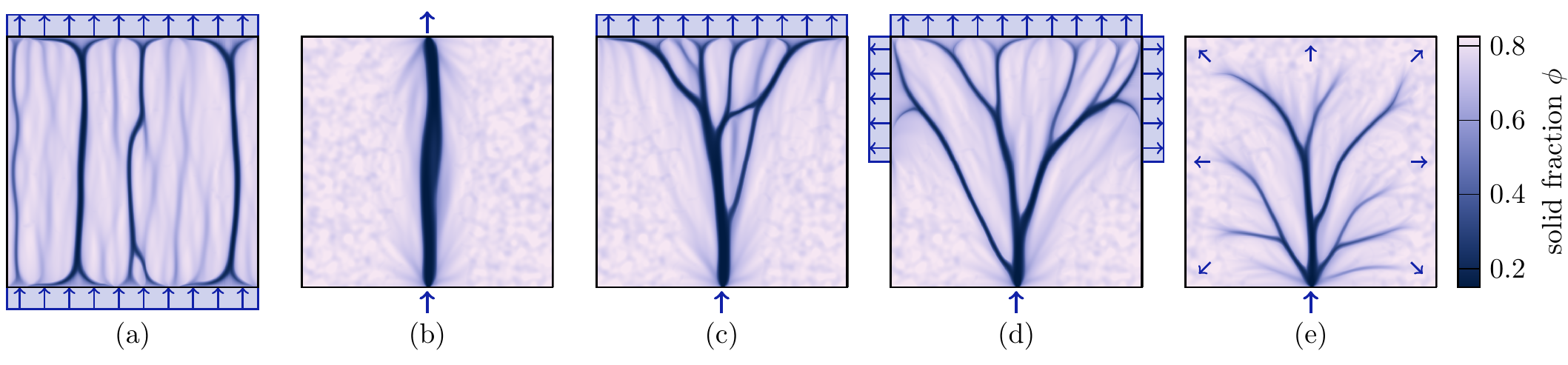}
	\caption{(color online) Erosion patterns as functions of boundary conditions obtained by solving \eqref{eq:p} and \eqref{eq:erode}. The solid fraction $\phi(\vx,t)$ is shown at $t=50$. The integrated flux through the system is ramped from zero to a final magnitude $\smash{F = \hat q_\IN w_\IN}$ over a duration $T=10$. Flow enters on the bottom wall and exits through the top wall (a--d) or via evaporation in the bulk (e). If the regions of inflow and outflow are of similar size as in (a) and (b), flow is concentrated in straight channels. If they are of different sizes as in (c--e), one inflow channel branches into many at the outflow.  The boundary widths satisfy $w_\IN = 0.1$ or 10; $w_\OUT = 0.1$, 10 or 20. Simulation parameters: grid size $1000^2$, $\phi_0 =0.8$, $\varphi_*=0.7$,  $\sigma_\phi=0.02$, $\zeta = 0.1$, $\xi = 0.1$, $\omega = 6.5$, $F=0.8$.}	
	\label{fig:branchTopology}
\end{figure*}

Equations \eqref{eq:p} and \eqref{eq:erode} together determine the evolution of the permeability of the porous medium, $\phi(\vx,t)$, and the pressure, $p(\vx,t)$, once we specify an initial condition.  Ignoring anisotropy in grain orientation and packing, we set \mbox{$\phi(\vx,0) =\phi_0 + \delta\phi(\vx)$}, with $\phi_0$ a constant and $\delta \phi$ a perturbed packing structure described as a random Gaussian thermal noise field with zero mean, variance $\sigma_\phi^2$, and correlation length $\zeta \gg l$, such that
\begin{equation}
	\left<\delta\phi(\vx)\delta\phi(\vy)\right>_r = \sigma_\phi^2 \exp\left[-\frac{r}{\zeta}\right].
\end{equation}
Here, $\left<*\right>_r= \int_\Omega (*)  d\vx d\vy / \text{vol}(\Omega)$ is a spatial average over all $\vx,\vy \in \Omega$ such that $|\vx-\vy|=r$.

The ratio of the correlation length to the stress communication length $\zeta/\xi$  controls the characteristic channel width $w_c$. From \eqref{eq:erode}, loss of solid material at a point reduces resistance to further erosion in a surrounding neighborhood of size $\xi$--- qualitatively similar to descriptions of nonlocal damage accumulation in settings such as hydraulic fracturing \cite{barenblatt2006}. Features in the $\phi$-field, initially of size $\zeta$, correspond to smoothed features in the $\varphi$-field. Thus, the channel width scaling satisfies $\smash{\zeta^2 < w_c^2 < \xi^2 + \zeta^2}$, approaching the small limit for large values of the packing variance $\sigma_\phi^2$ and vice versa, consistent with results obtained using the three-phase model \cite{mahadevan2012}. The width of a given channel scales with $w_c$ and varies with the amount of flux it conducts. See SI.4 for details.

Before considering the spatiotemporal evolution of the flow and permeability, we examine the local dependence of erosion on the threshold shape $\psi(\varphi)$ and local flux $\vq$. Letting $\smash{\left<*\right> = \int_A(*)d\vx / \text{vol}(A)}$ denote a spatial average over a mesoscopic region $A$, we introduce the scalar fields \mbox{$\Phi = \left<\phi\right>$}, \mbox{$G^2 = \left<|\gr p|^2\right>$}, and \mbox{$Q = \left<|\vq|\right>$}. We see they satisfy $Q = -\kappa(\Phi) G$, derived by averaging \eqref{eq:darcy}. Differentiating this relation and combining it with an averaged \eqref{eq:erode} yields a set of purely time-dependent equations describing trajectories through forcing--response phase space. For eroding states with $G^2 > \psi(\Phi)$,  
\begin{subequations}
\begin{equation}
	\frac{d\Phi}{dt} =  -\Phi \big(G^2 - \psi\big), \label{eq:homog_erode}
\end{equation}
	\vspace{-1.5em}
\begin{equation}
	\frac{d(G^2)}{dt} = 2 G^2 \left[\left(\frac{\Phi\kappa'}{\kappa}\right) \big(G^2 -\psi\big) + \frac{\dot Q}{Q}\right].
	\label{eq:G_partial}
\end{equation}\label{eq:homogSys}\end{subequations}
	
	Sustained erosion does not occur if $\dot Q = 0$, in which case points on the threshold surface \mbox{$G^2 = \psi$} are stable equilibria of the system. Eroding states reach the threshold in finite time, as can be seen from \eqref{eq:homog_erode}. For $\dot Q /Q > 0$, this is not the case. The quantity \smash{$\Phi\kappa'/\kappa <0$} is negative, so the squared-gradient decays or grows when the first or second term in \eqref{eq:G_partial} respectively dominates the other. The majority of the system's evolution takes place 
	along a monotonically increasing slow manifold $G_{\s}(\Phi)$ where the two are balanced, corresponding to $d(G^2)/dt = 0$, whence from \eqref{eq:G_partial} 
\begin{equation}
	G_{\s}(G_{\s}^2 - \psi) = -\frac{\dot Q}{\Phi \kappa'}, \label{eq:slow_man}
\end{equation}
	a cubic with one real root. In \figref{fig:homogSys}(a) we show the trajectories and slow manifolds for varying $\dot Q$. In \figref{fig:homogSys}(b) we plot the rate of erosion on the manifold, \mbox{$f_{\s} = G^2_{\s} - \psi$},  for thresholds of varying steepness at a particular flux rate \smash{$\dot Q$}. Theoretical bounds \mbox{$f_0 > f_\s > f_1$}, corresponding to constant thresholds $\psi = 0$ and 1, are plotted as dotted black lines. Both are monotonically increasing, diverge as $\Phi \to 1$, and vanish as $\Phi \to 0$, so the rate of erosion slows over long times. This effect is mitigated by a transition from $f\approx f_1$ to  $f_0$ near $\Phi \approx \varphi_*$. For sharp thresholds of large $\omega$, this effect is dominant and erosion accelerates upon reaching the transition region. The relative difference between the bounding rates, $(f_0-f_1)/f_1$, vanishes as $\Phi$ grows. It can be shown that for $\varphi_* > 1/2$, sharper thresholds yield faster average erosion over the entirety of the system's evolution. See SI.5 for details.

\begin{figure}
	\centering
	\includegraphics[width=\linewidth]{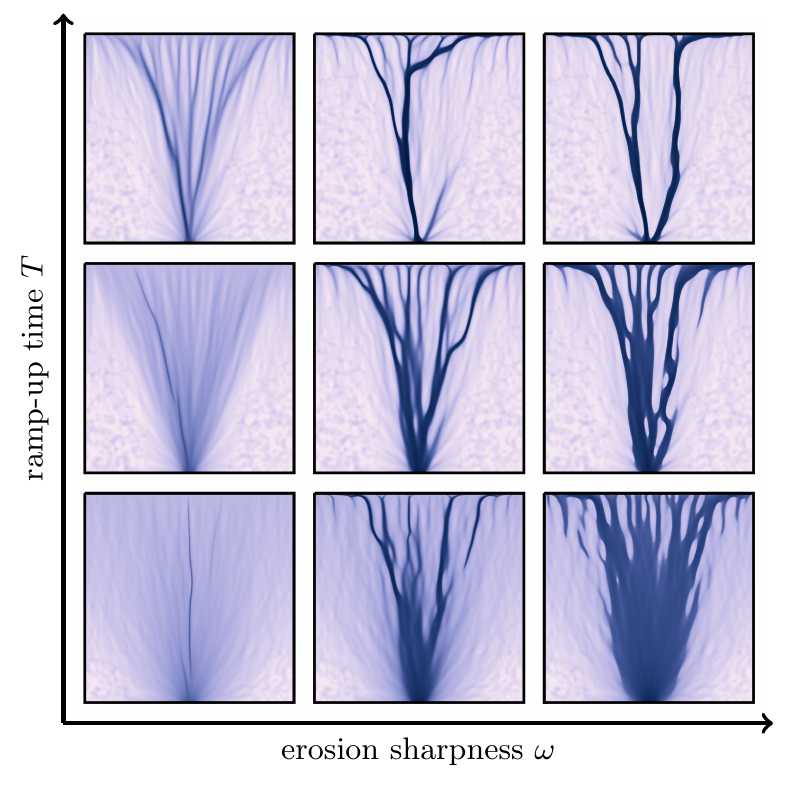}
	\caption{(Color online) Erosion patterns as functions of flux dynamics and threshold shape. The solid fraction $\phi(\vx,t)$ is plotted with the \figref{fig:branchTopology} color scheme at $\smash{t=50}$, subject to varied ramp time $T$ and threshold sharpness $\omega$. Increasing each parameter promotes confinement of erosion to a footprint which is smaller for more slowly increasing fluxes and larger for sharper thresholds. Boundary fluxes: $\smash{w_\IN=0.5}$, $\smash{w_\OUT=10}$, \smash{$\hat{s}=0$}, \smash{$F=0.5$}. Sharpness and ramp duration: \mbox{$\omega = \{1,8,15\}$}, \mbox{$T = \{0,3,10\}$}. Other parameters: \mbox{grid size 1024$^2$}, $\smash{\phi_0  = 0.8}$, $\smash{\varphi_* = 0.7}$, $\smash{\sigma_\phi = 0.02}$, $\smash{\zeta = 0.08}$, $\smash{\xi = 0.05}$.}
	\label{fig:phaseSpace}
\end{figure}

\paragraph{Branching morphospaces:}
We now turn to the spatiotemporal evolution of the flow  and  permeability fields in two-dimensional simulations. We aim to understand when, how and what arborization motifs arise  as a function of the boundary conditions, the dynamical rate of boundary fluxes, and the nature of the fragility/breaking threshold function. We integrate the coupled set of equations \eqref{eq:p} and \eqref{eq:erode} on a square domain \mbox{$\Omega=[-5,5]^2 \in \mathbb{R}^2$}, employing a second-order forward Euler method with Richardson extrapolation for error estimation and adaptive time stepping~\cite{heath1997}. See SI.6 for details. We adopt boundary conditions which ramp up the flux from zero over a duration $T$. Introducing \mbox{$r(t) = \min\left\{1,t/T\right\}$}, we set the fluid depletion rate and boundary fluxes as $s(\vx,t) = \hat s \ r(t)$, $q_\IN(\vx,t) = \hat q_\IN \ r(t), \hspace{1em} q_\OUT(\vx,t) = \hat q_\OUT \ r(t)$, where we have introduced a set of hatted constants corresponding to final magnitudes.
	This formulation yields a uniform bulk fluid sink $\hat s$ evenly distributed throughout the domain. Similarly, the boundary fluxes are assumed to be uniform everywhere on the regions  $\p\Omega_\IN$ and $\p \Omega_\OUT$, which we center on the bottom and top walls of the domain, respectively. We note the sign of $\hat s$ may be reversed and the labels ``in'' and ``out''  swapped with no change to morphogenic pattern formation, because the erosion rule \eqref{eq:erode_gen} is agnostic to the substitution $\gr p \to -\gr p$.

There are two feedback mechanisms through which erosion in the model promotes itself. The first, observed in the homogeneous system, is the threshold reduction due to previous erosion. The second is a direct effect of the coupling between flux and permeability. According to \eqref{eq:darcy}, flux is preferentially directed along paths of larger permeability, so that as it grows, flow from other parts of the domain is redistributed to eroded areas. In terms of the homogeneous phase space shown in \figref{fig:homogSys}(a), the resulting flux increase moves quickly eroding areas onto slow manifolds \smash{$G_s^2$} of higher $\smash{\dot Q}$, speeding up erosion. Slowly eroding areas experience the opposite effect until so much flow is diverted that \smash{$\dot Q \le 0$}, so erosion ceases. In this way, flow enhancement leads directly to selective erosion of high-$\kappa$ channelized regions of width $w_c$. For a given integrated fluid flux at the boundary $F$, the number of channels to form in the absence of geometric constraints will scale as $N_c \sim F/ w_c$; in what follows, $N_c > 1$.

	In \figref{fig:branchTopology} we show the results of simulations with four different combinations of boundary conditions and the role of bulk evaporation. In the first four panels \figref{fig:branchTopology}\mbox{(a--d)} we set $\hat s=0$ and consider the effect of variation in boundary flux width. Generically, if both the inlet width $w_\IN$ and outlet width $w_\OUT$ are larger than the emergent channel size $w_c$, boundary fluxes induce the formation of multiple channels, as seen in \figref{fig:branchTopology}(a). If both are less than $w_c$, a single channel is favored as in \figref{fig:branchTopology}(b). (We note branching in these settings is possible---\figref{fig:microMacro}(a) shows a single channel split and consolidate---but only given conveniently located low-$\kappa$ regions of the initial condition in the $\xi \ll \zeta$ limit.) If the reverse is true, i.e.  $w_\IN < w_c < w_\OUT$, then $N_c$ channels are created at the outlet and one at the inlet, as in  \figref{fig:branchTopology}(c) and (d). In \figref{fig:branchTopology}(e), we show the effect of bulk-evaporation driven flow with $\hat s>0$, a single inlet and no outlet. Because the channel width $w_c \ll L$ the system size, multiple channels form in the bulk, although their number and width is attenuated with distance from the inlet. These results may be summarized via a simple geometric argument suggesting a formula for reliable branch generation. If the number of channel heads distributed along the inlet and outlet are not the same, branching junctions arise in their linking, which is favored by flow continuity.
	
Finally, we consider the effects of varying the form of the erosion threshold function $\psi(\varphi)$ via its steepness $\omega$, and the rate of flux increase, via the ramp-up time $T$. \figref{fig:phaseSpace} shows a grid of eroded patterns corresponding to combinations of these two parameters. Low rates of flux increase correspond to slow manifolds $G_s^2$ close to the threshold $\psi$, so small drops in the pressure gradient can yield $|\gr p |^2 < \psi$. Conversely, rapidly increasing fluxes induce large pressure gradients $|\gr p|^2 \gg \psi$ before flow reorganization can occur, leading to large-scale washout,  also seen in the three-phase model~\cite{mahadevan2012}. We conclude that $T \gg 1$ is necessary for selective erosion. Increasing $\omega$ yields more erosion across the domain and appears to form sharper channel boundaries; this is consistent with the relationship between the erosion rate \mbox{$f=|\gr p|^2 - \psi$} and $\omega$ as in \figref{fig:homogSys}(b). As discussed, sharper thresholds induce faster average erosion, so more solid is eroded overall. In particular, systems with sharper thresholds have relatively higher rates of erosion in the region $\varphi<\varphi_*$. This speeds up flow enhancement and thus increases erosion selectivity.

	\paragraph{Conclusions:} Our minimal continuum model for the coupled dynamics of erosion, flow and permeability in a porous material shows how complex branching patterns can arise from simple causes.  While the model and discussion are rooted in the language of  fragile solids, our framework is broadly applicable beyond this setting, to branching patterns generated by local interactions subject to non-local flow constraints. Generalizing this to biological settings that feature non-linear couplings such as that between nutrient concentration and flow behavior, e.g. if portions of solid may be flow-seeking or flow-avoiding \cite{ocko2015} is a natural next step. 
	
	\paragraph{Acknowledgments:} C.H.R. and N.D. were partially supported by the National Science Foundation under Grant No.~DMS-1753203. C.H.R. was partially supported by the Applied Mathematics Program of the U.S. DOE Office of Science Advanced Scientific Computing Research under contract number DE-AC02-05CH11231. N.D. was partially supported by the NSF-Simons Center for Mathematical and Statistical Analysis of Biology at Harvard, award number 1764269, and the Harvard Quantitative Biology Initiative. L.M. was partially supported by the National Science Foundation under Grant Nos.~\mbox{DMR-2011754} and DMR-1922321.

\bibliographystyle{apsrev4-1}


%

\end{document}